\documentstyle[12pt]{article}
\begin{document}
\baselineskip 1.6\baselineskip
\begin{flushright}
{\bf KAIST-TPP-K20}
\end{flushright}
\vspace{0.3cm}
\begin{center}
{\bf {\Large The Dynamical Behaviors in (2+1)-Dimensional Gross-Neveu
Model with a Thirring Interaction}}
\end{center}
\vspace{0.3cm}
\begin{center}
Tae Seong Kim\footnote{e-mail: tskim@chep6.kaist.ac.kr},
Won-Ho Kye\footnote{e-mail: whkye@chep6.kaist.ac.kr},
and Jae Kwan Kim\\
\vspace{0.3cm}
{ \em  Department of Physics, \\ Korea Advanced Institute of Science and
  Technology, \\  373-1, Kusung-dong, Yusung-gu, Taejon, Korea }
 \end{center}

\vspace{0.3cm}
\begin{flushleft}
Abstract
\end{flushleft}

We analyze (2+1)-dimensional Gross-Neveu model with
a Thirring interaction, where
a vector-vector type four-fermi interaction
is on equal terms with a scalar-scalar type one.
The Dyson-Schwinger equation for fermion
self-energy function is constructed
up to next-to-leading order in 1/N expansion.
We determine the critical surface which is
the boundary between a broken phase and an unbroken one
in ($\alpha_c,~ \beta_c,~ N_c$) space.
It is observed that the critical behavior
is mainly controlled by Gross-Neveu coupling $\alpha_c$
and the region of the broken phase is separated
into two parts by
the line $\alpha_c=\alpha_c^*(=\frac{8}{\pi^2})$.
The mass function is strongly dependent upon
the flavor number N for $\alpha > \alpha_c^*$,
while weakly for $\alpha < \alpha_c^*$.
For $\alpha > \alpha_c^*$, the critical flavor number $N_c$
increases as Thirring coupling $\beta$ decreases.
By driving the CJT effective potential,
we show that the broken phase is energetically
preferred to the symmetric one.
We discuss the gauge dependence of the mass function and
the ultra-violet property of the composite operators.

\newpage
\section{Introduction}
Dynamical symmetry-breaking (DSB) plays an important role in
applying gauge field theories to particle physics.
The structure of chiral symmetry has
been studied extensively for
variety of gauge models [1 - 4].
Since the work of Nambu and Jona-Lasinio [1],
DSB has been investigated as the mechanism
of generating fermion
masses in elementary particle physics for
almost one generation [2 - 4].
Appelquist et al. [4] showed that
(2+1)-dimensional QED exhibits
a critical behavior as the flavor number N approachs
N$_c=32/\pi^2$ in the framework of the 1/N expansion.
Such a behavior was also confirmed by the lattice
simulation [5].

The search for the novel solution of Dyson-Schwinger (DS)
equation for a fermion propagator was initiated by Johnson,
Baker, and Willey [2]. The same equation can be derived
by extremizing the effective potential of
Cornwall, Jackiw, and Tomboulis (CJT) [6].
The CJT effective potential enables
one to check the stability of non-trivial solutions of
DS equation, that is, whether the solutions
are energetically preferred
to the trivial solution or not.

(2+1)-dimensional four-fermi interaction models
were shown to be renormalizable in the framework of the
1/N expansion despite its non-renormalizability
in ordinary weak coupling expansion [7].
Such a dramatic transmutation is due to the fact
that the composite operator acquires
the large anomalous dimension in strongly-correlated region.
(2+1)-dimensional Gross-Neveu (GN) model possesses a non-trivial
ultra-violet (UV) fixed point at the leading order of
1/N expansion, which survives beyond the leading order [8].
It provides a ground to study DSB.
(2+1)-dimensional Thirring model
has also been studied in the context of the gauge structure
and a dynamical mass generation
of the fermions in the 1/N expansion [9].
It is reported that Thirring model shows the phase transition
with the strong dependence of the critical flavor number
on Thirring coupling constant, and possesses
non-trivial UV fixed point albeit at non-perturbative order in 1/N [9].
While the gap equation at the leading order is local
in GN model, the first non-trivial gap equation is non-local
in Thirring model.
It turns out that the gap equation is a relation
to fine-tune the coupling as in GN model rather than
the value of the fermion mass as in
(2+1)-dimensional QED.
The dynamically generated fermion mass then becomes a physical
parameter which is much smaller than the natural cut-off of the
theory.

Thirring interaction is an another independent
composite operator which is relevant
to GN interaction in strongly coupled regime [8, 9].
Accordingly, it is natural for us to consider
whether the above four-fermi interactions co-operate dynamically
in the generation of fermion mass.

In this paper, we consider (2+1)-dimensional
Gross-Neveu model with
a Thirring interaction, and investigate how Thirring
interaction plays a part on the critical behavior
of GN model .
In Sec. 2, we introduce our model and derive the dressed
propagators whose UV behavior improved by
adding the gauge-fixing-like term to the Lagrangian.
In Sec. 3, the DS equation is constructed up to 1/N-order.
We determine the critical surface
which is the boundary between
a broken phase and an unbroken
one in ($\alpha, \beta, N$) space,
and also present the fermion mass function
in an appropriate approximation scheme.
The critical behavior
is mainly controlled by the Gross-Neveu coupling $\alpha_c$
and the region of the broken phase is separated
into two parts by
the line $\alpha_c=\alpha_c^*(=\frac{8}{\pi^2})$.
The gauge dependence of the mass function is discussed.
In Sec. 4, we show that our nontrivial solution
is energetically preferred to the trivial one by
deriving the CJT effective potential.
In Sec. 5, we discuss
the consequences of results and the UV
property of the composite operators.

\section{The Model}
Our model is given in the Euclidean version by the Lagrangian
\begin{eqnarray}
{\cal L}=\bar{\psi}(i\gamma \cdot \partial)\psi +\frac{1}{2N}
(~~g^2(\bar{\psi}\psi)^2+
h^2(\bar{\psi}\gamma_\mu\psi)^2~~),
\end{eqnarray}
where $\psi$ are the four-component Dirac spinors
whose flavor indices are suppressed and  N is the
fermion flavor number.
As one can see, scalar-scalar and vector-vector
type four-fermion interactions
are parallelly introduced in
the above equation.
The $\gamma$ matrices
are chosen to be anti-hermitean
as follows,
\begin{eqnarray}
\gamma_0=\left( \begin{array}{cc}
                0  & i \mbox{I}\\
                i \mbox{I}      &0
                \end{array}
                \right),~~
\gamma_j=\left( \begin{array}{cc}
                0  & -\sigma_j\\
                \sigma_j       & 0
                \end{array}
                \right),
\end{eqnarray}
where I is $2\times 2$ identity matrix
, $\sigma_j$'s are the Pauli matrices and $j=1, 2$.

Introducing the auxiliary scalar field $\sigma$
and the auxiliary vector field $A_\mu$ to
facilitate 1/N expansion [7, 9], we can rewrite Eq. (1) as
\begin{eqnarray}
{\cal L}^{eq}=\bar{\psi}(i\gamma \cdot \partial)
\psi- \frac{1}{2h^2}A_\mu^2
-\frac{1}{2g^2}\sigma^2 +
\frac{1}{\sqrt{N}}A_\mu (\bar{\psi}\gamma_\mu\psi)
+\frac{1}{\sqrt{N}}\sigma(\bar{\psi}\psi).
\end{eqnarray}
It is easy to see the equivalence between ${\cal L}$ and ${\cal L}^{eq}$
as follows
\begin{equation}
{\cal L}=-\ln\left\{\int {\cal D}\sigma {\cal D}A_\mu
{}~\exp[-\int {\cal L}^{eq} d^3x] \right\}.
\end{equation}
If we do not want to worry about what quantities should be required
to be renormalizable, we can improve the UV behavior of
all the Green's functions by adding to Eq. (3) the term
$(\partial_\mu A_\mu)^2 /(2\xi)$  and postulating that the observables
are those quantities which are independent of the gauge
parameter $\xi$.
The new theory also has a restricted gauge symmetry [9].

The tree propagators, after the introduction of
the gauge-fixing term, are
\begin{eqnarray}
\Delta^{(0)}_{\mu\nu}(p)&=&h^2(\delta_{\mu\nu}-p_\mu p_\nu/p^2)
                         +\frac{\xi}{p^2+\xi/h^2}(p_\mu p_\nu/p^2),\\
G^{(0)}(p)&=&g^2.
\end{eqnarray}
To find the 1/N leading propagator,
we evaluate the three kinds of one-loop
two-point functions by the method of dimensional
regularization as follows:
\begin{eqnarray}
\Pi_{\mu\nu}(p)&=&-\mbox{tr}\int\frac{d^3k}{(2\pi)^3}\gamma_\mu
       \frac{1}{\gamma \cdot k +m}\gamma_\nu
       \frac{1}{\gamma \cdot (k-p)+m}\nonumber\\
       &=& -(\delta_{\mu\nu}-p_\mu p_\nu/p^2)F(p), \\
\Pi(p)&=&-\mbox{tr}\int\frac{d^3k}{(2\pi)^3}
        \frac{1}{\gamma \cdot k +m}
        \frac{1}{\gamma \cdot (k-p)+m}\nonumber\\
       &=& -2 F(p),\\
\Pi_\mu(p)&=&-\mbox{tr}\int\frac{d^3k}{(2\pi)^3}
       \frac{1}{\gamma \cdot k +m}\gamma_\mu
       \frac{1}{\gamma \cdot (k-p)+m}\nonumber\\
       &=& 0,
\end{eqnarray}
where the trace is acting on a spinor space , $m$ is the
fermion mass generated
dynamically through 1/N corrections and
$F(p)$ can be expressed without introducing
the regulating parameter as follows [3, 4]
\begin{eqnarray}
F(p)=\frac{1}{\pi}\left[\frac{m}{2}
     +\frac{p^2-4m^2}{4\sqrt{p^2}}\arcsin
     (\sqrt{\frac{p^2}{p^2+4m^2}})\right].
\end{eqnarray}
We adapt the approximation
$m^2 \ll p^2$ in the following analysis and in that case
the $F(p)$ goes to $\sqrt{p^2}/8$
for the evaluation of DS gap equation in Sec. 3.
There is no mixing
between the scalar field $\sigma$ and the
vector field $A_\mu$, since
the scalar-vector one-loop diagram vanishes (Eq. (9)).

We can evaluate the dressed propagators by summing the chain
diagrams as follows:
\begin{eqnarray}
\Delta_{\mu\nu}(p)&=&\frac{1}{1/h^2+F(p)}
                    (\delta_{\mu\nu}-p_\mu p_\nu/p^2)+
                    \frac{\xi}{p^2+\xi/h^2}
                    (p_\mu p_\nu/p^2),\\
G(p)&=&\frac{1}{1/g^2+2F(p)}.
\end{eqnarray}

\section{The Dyson-Schwinger Equation}
We start with the theory with a UV cut-off
as was usually done in QED$_{3+1}$ [10].
By using the dressed propagators, we can construct
the following DS gap equation [2 - 4] up to 1/N-order,
\begin{eqnarray}
S^{-1}(p)&=& S^{(0)~-1}(p)+{\sqrt{N}}G(0)\int ^\Lambda_0
           \frac{d^3k}{(2\pi)^3}\Gamma(k)\mbox{tr}[S(k)]\nonumber\\
         & &- \frac{1}{N} \int ^\Lambda_0
         \frac{d^3 k}{(2\pi)^3}\gamma_\mu
           S(k) \gamma_\nu \Delta_{\mu\nu}(p-k)
 -\frac{1}{N} \int ^\Lambda_0 \frac{d^3 k}{(2\pi)^3}
           S(k)  G(p-k),\nonumber\\
\end{eqnarray}
where
\begin{eqnarray}
  S^{-1}(p)&=&A(p)\gamma \cdot p +\Sigma(p),\\
  S^{(0)~-1}(p)&=&\gamma \cdot p ,\\
  \Gamma(p)&=&\frac{1}{\sqrt{N}}(\frac{\Lambda^2}{p^2})^\gamma,~~
  \gamma=\frac{3}{\pi^2N}, \\
  A(p)&=&(\frac{\Lambda^2}{p^2})^{\gamma_\psi}, ~~
  \gamma_\psi=\frac{-1}{\pi^2N}.
\end{eqnarray}

$\Sigma(p)$ is the fermion self-energy function generated dynamically
and $\Lambda$ is the natural cut-off of the theory.
$A(p)$ and $\Gamma(p)$ denote the renormalization constants
of the wave function and the vertex, respectively.
They can be calculated perturbatively from the bare Lagrangian
${\cal L}^{eq}$ and their detailed derivations
are presented in the Appendix.
The second term in the right hand side of Eq. (13) denotes the
tadpole diagram.
We can not  neglect the corrections of 1/N-order
which stems from $A(p)$ and $\Gamma(p)$,
since we want to consider the theory up to that order.
The extra two terms denote the self-energy diagrams by
vector field $A_{\mu}$ and scalar field $\sigma$.
The crossing diagram
between $\Delta_{\mu\nu}(p)$ and $G(p)$ correspond to the
order of $1/N^2$, therefore, it is excluded from the above equations.

Taking the trace over the gamma matrices in Eq. (13),
we get the following DS gap equation
\begin{eqnarray}
\Sigma(p) &=&4g^2\int ^\Lambda_0 \frac{d^3k}{(2\pi^3)}
             \frac{\Sigma(k)}{k^2+\Sigma^2(k)}\nonumber\\
          & &+4g^2\int ^\Lambda_0 \frac{d^3k}{(2\pi^3)}
             \frac{\Sigma(k)}{k^2+\Sigma^2(k)}\ln(\Lambda^2/k^2)
             \left[\gamma-2\gamma_\psi \frac{k^2}{k^2+\Sigma^2(k)}
             \right]\nonumber\\
          & &+\frac{2}{N}\int ^\Lambda_0 \frac{d^3k}{(2\pi^3)}
             \frac{\Sigma(k)}{k^2+\Sigma^2(k)}
             \frac{1}{|p-k|/8+1/h^2}\nonumber\\
 & & -\frac{1}{N}\int ^\Lambda_0 \frac{d^3k}{(2\pi)^3}
           \frac{\Sigma(k)}{k^2+\Sigma^2(k)}
           \frac{1}{|p-k|/4+1/g^2}\nonumber\\
& & +\frac{\xi}{N}\int\frac{d^3k}{(2\pi)^3}
\frac{\Sigma(k)}{k^2+\Sigma^2(k)}
    \frac{1}{(k-p)^2+\xi/h^2},
\end{eqnarray}
where $|p-k| =\sqrt{(p-k)^2}$.

In the case of N$\leq$1 it is meaningless for us to analyze Eq. (18)
since the 1/N expansion is failed in that region.
We intend to present the comparable feature with the previous studies [4, 5],
in which it is reported that the fermion mass is vanished
when the fermion flavor N goes to N$_c=32/\pi^2$.
We plot the following figures in the case of N$=$1
for illustrative purposes.

Since the last term in Eq. (18) has
the gauge parameter $\xi$,
$\Sigma(p)$ is a gauge-dependent quantity.
However, the fact that
it is not identically zero has physical consequence
( i.e., chiral-symmetry breaking). Accordingly, whether the
fermion mass is generated or not is the gauge invariant
statement [9].

\subsection{Critical surface in ($\alpha_c, \beta_c, N_c$) space}
Let us approximate $\Sigma(k)$ in Eq. (18) to  $m \approx
\Sigma(k)|_{k=0}$.
Taking the limit $m\rightarrow 0$, we
can access the critical region as follows:
\begin{equation}
(g^2, h^2, N)\stackrel{m\rightarrow 0}{\longrightarrow}
(g_c^2, h_c^2, N_c).
\end{equation}
In that case, the critical surface is defined by
\begin{eqnarray}
1 &=& \frac{8}{\alpha_c\pi^2}\left(1+\frac{10}{N_c\pi^2}\right)
     +\frac{8}{N_c\pi^2}\ln\left(\frac{1+\beta_c}{\beta_c}\right)-
  \frac{2}{N_c\pi^2}\ln\left(\frac{1+\alpha_c}{\alpha_c}\right) \nonumber\\
  & & +\frac{4}{N_c\pi^2}\sqrt{\frac{\tilde{\xi}}{\beta_c}}
  \arctan\left(\frac{1}{\sqrt{\tilde{\xi} \beta_c}}\right),
\end{eqnarray}
where the dimensionless quantities are given by,
\begin{equation}
\alpha_c=\frac{4}{\Lambda g_c^2}, ~~\beta_c=\frac{8}{\Lambda h_c^2},
 ~~\tilde{\xi}=\frac{\xi}{8 \Lambda}.
\end{equation}

{}From Eq. (20), regardless of $\beta_c$,
there exists the critical flavor $N_c$ only
in the region of $\alpha_c > \alpha_c^*$
and $\alpha_c$ goes to $\alpha_c^*(=\frac{8}{\pi^2})$
as $N_c$ goes to infinity.
Eq. (20) is plotted in Fig. 1 and Fig. 2 for two
different gauge conditions, which are  $\xi=0$ (Landau gauge)
and $\xi=1$ (Feynman gauge), respectively.
Each curve in Fig. 1 and Fig. 2 corresponds to the boundary
between a broken phase and an unbroken one.
One can see that the curves of $\beta_c = 1.0 \times 10^{4}$ are
not deformed in two figures.
The other curves in Fig. 2 are shifted to the right for the
corresponding curves in Fig. 1.

\subsection{Fermion mass function}
After taking the integrations in Eq. (18), we obtain the following
equation:
\begin{equation}
N=\frac{\frac{8}{\alpha\pi^2}B(M)+
\frac{1}{\pi^2}C(M)+\frac{4 \tilde{\xi}}{\pi^2}D(M)
}{1-\frac{8}{\alpha\pi^2}A(M)},
\end{equation}
where
\begin{eqnarray}
M&=&m/\Lambda\\
A(M)&=&1-M \arctan(1/M)\\
B(M)&=& \frac{1}{\pi^2}\left[
	 (-10a-5b-\frac{10}{3}c-2bM^2+14cM^2) \right. \nonumber\\
    & & \left. +(12a-16cM^2)M \arctan(1/M) \right. \nonumber\\
    & & \left. +7bM^2 \ln\left(\frac{1+M^2}{M^2}\right)
    -(a-bM^2-cM^2)\frac{M^2}{1+M^2}\right] \\
C(M)&=&8F(\beta, M)-2F(\alpha, M),\\
D(M)&=&\frac{4~ \tilde{\xi}}{\pi^2}~\frac{M \arctan(1/M) -
\sqrt{\tilde{\xi}\beta} \arctan(1/\sqrt{\tilde{\xi}\beta})}
{M^2 - \tilde{\xi}\beta},\\
F(\alpha, M)&=&\frac{1}{M^2+\alpha^2}
\left[\frac{M^2}{2}\ln\left(
\frac{1+M^2}{M^2}\right)-M\alpha\arctan(1/M)+
\alpha^2\ln\left(\frac{1+\alpha}{
\alpha}\right)\right], \nonumber\\
\end{eqnarray} where
we have taken the following interpolation:
\begin{eqnarray}
& x\ln x \approx a x+ bx^2+cx^3, ~~\mbox{for}~~x \in [0, 1]\\
& \mbox{with}~ a= -2.5,~ b=4.0,~ c=-1.5. \nonumber
\end{eqnarray}
The fermion mass functions in Landau gauge
are presented for  $\alpha < \alpha_c^*$ and
$\alpha > \alpha_c^*$ in Fig. 3 and Fig. 4, respectively.
The fermion mass functions in Feynman gauge
are presented for  $\alpha < \alpha_c^*$ and
$\alpha > \alpha_c^*$ in Fig. 5 and Fig. 6, respectively.
The mass functions in Fig. 3 and Fig. 5 are strongly dependent on
the flavor number N, while those in Fig. 4 and Fig. 6 weakly.
One can easily confirm that the Eq. (22) is equivalent to Eq. (20)
in the limit of $M \rightarrow 0$.

\section{The Vacuum Stability}
The DS equation is obtained by extremizing the CJT [5] effective
potential with respect to fermion full propagator $\bar{S}(p)$.
The CJT effective potential for our model is given by
\begin{eqnarray}
V^{CJT}(\bar{S})&=& \int \frac{d^3q}{(2\pi)^3} \mbox{tr}\left[
            \ln(S^{(0)~-1}(q)\bar{S}(q))-S^{(0)~-1}(q)\bar{S}(q)+1\right]
	    \nonumber\\
          & & -\frac{\sqrt{N}}{2}G(0)\left(\int\frac{d^3q}
          {(2\pi)^3}\Gamma(q)\mbox{tr}
             [\bar{S}(q)]\right)^2\nonumber\\
          & & +\frac{1}{2N}\int\frac{d^3qd^3k}{(2\pi)^6} \mbox{tr}
          \left[\gamma_\mu \bar{S}(q)\gamma_\nu \bar{S}(q+k)\Delta_{\mu\nu}(k)
          +\bar{S}(q)\bar{S}(q+k)G(k) \right],\nonumber\\
\end{eqnarray}
where the first term in the right hand side of Eq. (30)
denotes the fermion one-loop contribution
and the extra terms denote the two-loop contributions.
We consider the extremal condition in the CJT effective potential
as follows
\begin{eqnarray}
\frac{\delta V^{CJT}(\bar{S})}{\delta \bar{S}(p)}|_{\bar{S}=S}=0.
\end{eqnarray}
In that case, we meet the following equation:
\begin{eqnarray}
S(p)^{-1}=S^{(0)~-1}(p)+\Xi(p),
\end{eqnarray}
where,
\begin{eqnarray}
\Xi(p)&=&\sqrt{N}G(0)\int\frac{d^3k}{(2\pi)^3}\Gamma(k)
             \mbox{tr}[S(k)]\nonumber\\
         & &-\frac{1}{N} \int \frac{d^3 k}{(2\pi)^3}\gamma_\mu
           S(k) \gamma_\nu \Delta_{\mu\nu}(p-k)
 -\frac{1}{N} \int \frac{d^3 k}{(2\pi)^3}
           S(k)G(p-k). \nonumber\\
\end{eqnarray}
One see that Eq. (32) with Eq. (33)
coincides with the previous
DS equation presented in Eq. (13).
By inserting the Eq. (32) with Eq. (33)
into the Eq. (30), we obtain the CJT effective potential
at the extremal propagator $S(p)$ as follows
\begin{eqnarray}
V^{CJT}(S)=-2\int \frac{d^3p}{(2\pi)^3}\left[\ln(1+
                 \frac{\Sigma(p)^2}{p^2})
                 -\frac{\Sigma^2(p)}{p^2+\Sigma^2(p)}
                 \right].
\end{eqnarray}
Since the function $\ln(1+x)-x/(1+x)$ is positive for all positive $x$,
then $V^{CJT}(S)$ is less than or equal to zero.
Hence, the energy of any nontrivial solutions is lower than that of
the trivial (perturbative) solution $\Sigma(p)=0$, therefore, the broken
phase is always energetically preferred to
the symmetric one.

\section{Discussion}
We studied (2+1)-dimensional Gross-Neveu model
with a Thirring interaction,
where a vector-vector type four-fermi interaction
is on equal terms with a scalar-scalar type one.
To solve the DS gap equation up to 1/N-order, the
renormalization constants ( {\it i.e.}, $A(p),~ \Gamma(p)$)
were calculated perturbatively up to that order and
the fermion self-energy function $\Sigma(p)$ was approximated to
the constant $m (\approx \Sigma(p=0))$.
We expect that these approximations do not change
the qualitative feature of the phase structure of our model.

Our gap equation (Eq. (18)) contains
a gauge parameter $\xi$, and thus it is gauge dependent.
The position $A(p)^2p^2=-\Sigma(p)^2$ of the pole of fermion propagator
is a physical quantity that must
be independent upon the parameter.
We, unfortunately, have ignored the momentum dependence of
$\Sigma(p)$ in our analysis, we could not present the  physical
mass of fermion.
The numerical evaluation of Eq. (18) will allow one
to investigate the physical mass of fermion.
However, as was discussed in Ref. [9], the fact that {\it the fermion
mass is not identically zero has
physical consequence} (i. e., chiral-symmetry breaking).
Accordingly, whether the fermion mass is generated
or not is a gauge invariant statement.
The critical surfaces for
$\tilde{\xi}=0.0$ (Landau gauge) and
$\tilde{\xi}=1.0$
(Feynman gauge) are presented in the contour shape
in Fig. 1 and Fig.2, respectively.
The curves of $\beta_c=1.0\times10^4$ are not deformed
in two gauge conditions, since
in that case our model is
equivalent to GN model
in the bare Lagrangian level.
The other curves in Fig. 2 are shifted
to the right from the corresponding curves in Fig. 1.
Such a feature is natural since it means that
the vector bosons are more correlated
than the scalar bosons in the neighborhood of
the phase boundary.

The critical line $\alpha_c=\alpha_c^*(=\frac{8}{\pi^2})$
in the Gross-Neveu model of the leading order turns into
a critical surface
in $(\alpha_c, \beta_c, N_c)$ space
in our model (see Fig. 1 and Fig. 2).
The fermion mass function in Landau gauge are presented
in the region of $\alpha < \alpha_c^*$ and $\alpha > \alpha_c^*$
in Fig. 3 and Fig. 4.
They show that the fermion mass is generated dynamically
in both regions.
The fermion mass is weakly dependent on the fermion flavor number $N$
in the former,
while strongly in the region of the latter.
Such features can be reasonably understood.
For $\alpha <\alpha_c^*$, the critical behaviors are dominated
by the tadpole diagram, then the theory always shows
the mass generation as to all of $N$.
However, for $\alpha > \alpha_c^*$, the tadpole
term does not contribute to the mass generation,
therefore the critical behaviors are controlled by the
terms of the order of $1/N$ and they are
QED$_{2+1}$-like or Thirring-like.
As argued in Ref. [9],
the effective infrared coupling weakens
like $1/N$  as $N$ increase
due to the screening effect of fermions.
Therefore, the infrared coupling
becomes so weak that the fermion condensates cannot occur
in the case of $N > N_c$.
All of the curves in Fig. 1 approaches to the line $\alpha_c^*$ as $N$
increase, since our model in the large $N$ limit is
the GN model of the leading order (Eq. (18) and Eq. (20)).
The fermion mass function in Landau gauge are also presented
in the region of $\alpha < \alpha_c^*$ and $\alpha > \alpha_c^*$
in Fig. 5 and Fig. 6.

It is interesting to discuss the UV property [7]
of composite
operators ( {\it i.e.}, $(\bar{\psi} \psi)^2$,
and $(\bar{\psi}\gamma_\mu \psi)^2$)
in the strongly correlated region.
{}From the dressed propagators (Eq. (11) and Eq. (12)) we can see the
asymptotic behavior
in UV region ({\it i.e.}, $p \rightarrow $
$\Lambda$) as follows:
\begin{eqnarray}
\Delta_{\mu\nu}(p) &\sim& \frac{1}{p}
(\delta_{\mu\nu}-p_\mu p_\nu/p^2), \nonumber \\
G(p) &\sim& \frac{1}{p}, \nonumber
\end{eqnarray}
thus the mass dimensions
of the fields in that
region are  $
[\sigma]_{UV}=1$, and $[A_\mu]_{UV}=1$.
Noting the equivalence $\sigma \sim \bar{\psi}\psi$ and
$A_\mu \sim \bar{\psi}\gamma_\mu \psi$
from the Lagrangian (Eq. (3)), we see the following  facts:
\begin{eqnarray}
\left[(\bar{\psi}\psi)^2\right]_{UV} &=& 2, \nonumber \\
\left[(\bar{\psi}\gamma_\mu \psi)^2 \right]_{UV} &=& 2. \nonumber
\end{eqnarray}
Accordingly, the renormalizability
in that region is guaranteed [7].

Our model contains two of the most fundamental types
of four-fermi interactions,
and they co-operate dynamically in the generation of fermion mass.
In the region of $\alpha < \alpha_c^*$,
the general features of mass generation is not
disturbed by the higher $1/N$-order corrections
by the dominant tadpole term.
In that of $\alpha > \alpha_c^*$, Thirring interaction
plays important roles not only in generating
the fermion mass dynamically but also in making its strong
dependence on N.
In the latter, the critical flavor number $N_c$
is increases as $\beta$ decreases.
Our model, therefore, is probably
a good ground for studying the fundamental phase transitions
of {\it the Nature}.
It is important to evaluate Eq. (18) numerically to check
the stability [9] of the results.
Authors are in process for analyzing it.

This work is supported in part by KOSEF (Korea Science
and Engineering Foundation).
\newpage
\section*{Appendix}
We derive the renormalization constants
of the wave function and the vertex up to 1/N-order.
The former can be calculated from the self energy diagrams
by the vector field and the scalar field,
\begin{eqnarray}
{\cal D}_1+{\cal D}_2&=&\frac{1}{N}\int
\frac{d^3k}{(2\pi)^3}\gamma_\mu \Delta(k)_{\mu\nu}\gamma_\nu
S^{(0)}(p-k)
 +\frac{1}{N}\int \frac{d^3k}{(2\pi)^3}G(k)S^{(0)}(p-k)\nonumber\\
&= & -\frac{4-1}{3N\pi^2}~(\gamma \cdot p)~
\ln (\Lambda^2/p^2)+ \mbox{finite terms}.
\end{eqnarray}
Thus the wave function renormalization constant is given by
\begin{equation}
A(p)=1 -\frac{1}{N\pi^2}\ln (\frac{\Lambda^2}{p^2}).
\end{equation}
The latter can be calculated by the vertex diagrams
by the vector field and  the scalar field,
\begin{eqnarray}
{\cal V}_1+{\cal V}_2&=&\frac{1}{N^{3/2}}\int
\frac{d^3k}{(2\pi)^3}\gamma_\mu
\Delta(k)_{\mu\nu}\gamma_\nu S^{(0)}(p_1+k)
S^{(0)}(p_2+k) \nonumber\\
& &+\frac{1}{N^{3/2}}\int
\frac{d^3k}{(2\pi)^3}G(k) S^{(0)}(p_1+k) S^{(0)}(p_2+k) \nonumber\\
&=& \frac{4-1}{N^{3/2}\pi^2}\ln(\Lambda^2/p^2_{max})
+\mbox{finite terms},
\end{eqnarray}
where $p_{max}^2$ is the largest of $p_1^2$
and $p_2^2$ with $p_1$ and $p_2$
the incoming and outgoing fermion momenta, respectively.
The vertex renormalization constant for tadpole
diagram is given by
\begin{eqnarray*}
\Gamma(p)&=&\frac{1}{\sqrt{N}}+\lim_{p_1, p_2 \rightarrow p}
\frac{3}{N^{3/2}\pi^2}\ln(\Lambda^2/p^2_{max}) \nonumber\\
 &=&\frac{1}{\sqrt{N}}(1+\frac{3}{N\pi^2}\ln(\Lambda^2/p^2)).
\end{eqnarray*}

\newpage
\section*{References}
\begin{description}
\item{[1]} Y. Nambu and G. Jona-Lasinio, Phys. Rev. {\bf  122}, 345 (1961).
\item{[2]} K. Johnson, M. Baker, and R. Willey, Phys. Rev. {\bf 136},
           B1111 (1964); Phys. Rev. {\bf 163}, 1699 (1967).
\item{[3]} R. D. Pisarski, Phys. Rev. {\bf D}29, 2423 (1989).
\item{[4]} T. Appelquist, M. J. Bowick, D. Karabali, and L. C. R.
           Wijewardhana, Phys. Rev. {\bf D 33}, 3774 (1986); T. Appelquist,
           D. Nash, and L. C. R. Wijewardhana, Phys.
           Rev. Lett.  {\bf 60}, 2575 (1988).
\item{[5]} E. Dagotto, A. Kocic and J. B. Kogut,Phys. Rev. Lett. {\bf 62},
           1083 (1989).
\item{[6]} J, M. Cornwall, R. Jackiw, and E. Tomboulis,
          Phys. Rev. {\bf D 10}, 2428 (1974).
\item{[7]} B. Rosenstein, B. J. Warr, and S. H. Park,
           Phys. Rev. Lett. {\bf 62}, 1433 (1989); {\it ibid.}, Phys. Rep.
           {\bf 205}, 59 (1991) and references therein.
\item{[8]} G. Gat, A. Kovner, and B. Rosenstein, Nucl. Phys. {\bf B } 385,
           (1992).
\item{[9]} M. Gomes, R. S. Mendes, R. F. Ribeiro, and A. J. da Silva, Phys.
           Rev. {\bf D 43}, 3516 (1991); D. K. Hong and S. H. Park, Preprint
           SNUTP 93-47.
\item{[10]} J.B.Kogut,E.Dagotto, and A.Kocic, Phys. Rev. Lett. 60 (1988)
           772; E.Dagotto and J. G. Kogut, Nucl. Phys. B295 (1988) 125.
\end{description}
\newpage
\section*{Figure Captions}
\begin{description}
\item{Fig. 1:} The projection of the critical surface
               on ($N_c, \alpha_c$) plane for various $\beta_c$.
         $\tilde{\xi}$ is chosen to be zero (Landau gauge).
\item{Fig. 2:} The projection of the critical surface
               on ($N_c, \alpha_c$) plane for various $\beta_c$.
         $\tilde{\xi}$ is chosen to be one (Feynman gauge).
\item{Fig. 3:} The fermion mass $M$ as a function of the flavor
	       number $N$ for various $\alpha$ ($\alpha <\alpha_c^*$)
	       and $\beta$ in $\tilde{\xi}=0$.
\item{Fig. 4:} The fermion mass $M$ as a function of the flavor
	       number $N$ for various $\alpha$ ($\alpha >\alpha_c^*$)
	       and $\beta$ in $\tilde{\xi}=0$.
\item{Fig. 5:} The fermion mass $M$ as a function of the flavor
	       number $N$ for various $\alpha$ ($\alpha <\alpha_c^*$)
	       and $\beta$ in $\tilde{\xi}=1$.
\item{Fig. 6:} The fermion mass $M$ as a function of the flavor
	       number $N$ for various $\alpha$ ($\alpha >\alpha_c^*$)
	       and $\beta$ in $\tilde{\xi}=1$.
\end{description}

\end{document}